# Two-phonon Raman bands of bilayer graphene: revisited

Valentin N. Popov*

Faculty of Physics, University of Sofia, BG-1164 Sofia, Bulgaria

**Abstract**

We present complete calculations of the two-phonon Raman bands of bilayer graphene, including all overtone and combination modes, within a density-functional tight-binding model. Based on our results, we assign unambiguously the observed two-phonon Raman bands to two-phonon modes, thus resolving the existing controversies. In particular, we show that both overtone and combination modes have essential contribution to the 2D band, bringing about specific modifications of the band shape. We argue that a mid-range two-phonon Raman band, previously assigned to the 2ZO mode, should be assigned to the TOZO' mode. We find that the Raman band, usually assigned to the LOLA mode, has significant contribution from the TOZO mode. The predicted Raman bands can be used for assignment of the observed ones in the Raman spectra of bilayer graphene for the needs of sample characterization for future technological applications.

*Corresponding Author. Fax: +359 2 96 25 276. Tel.: +359 2 81 61 478

*E-mail address*: vpopov@phys.uni-sofia.bg

# 1. Introduction

Bernal-stacked bilayer graphene (BLG) has unique electronic properties that make it a promising material for application in nanoelectronics [1,2]. Recently, an opening of a tunable band gap has been observed by application of an external electric field to BLG [3] and by molecular doping [4], which opens the way for the use of BLG in electronic devices. This progress has prompted further experimental and theoretical investigations of BLG [5]. A key point in the experimental work on BLG is the in-depth characterization of the synthesized samples. Raman spectroscopy has proved to be an inexpensive and nondestructive characterization method for carbon allotropes [6-8]. In this method, the assignment of the observed Raman bands is facilitated by theoretical predictions for the studied material.

The experimental Raman spectra of BLG exhibit two-phonon bands, similar to those observed and exhaustively studied in single-layer graphene (SLG) [8], and usually denoted as 2D, 2D', D + D'', etc., which are attributed to *resonant Raman scattering processes*. The Raman intensity of the two-phonon bands can be described in the framework of the quantum-mechanical perturbation theory [9]. The relevant terms of the perturbation series can be interpreted as processes of electron-hole creation and annihilation, and electron/hole scattering by phonons. These terms include matrix elements, describing interactions between photons, electrons/holes, and phonons. A geometrical approach, neglecting the interaction matrix elements, has allowed revealing the contributions to the 2D band of SLG from different resonant processes and reconstructing the phonon dispersion [10]. The dependence of the 2D band of BLG on the laser excitation has been calculated within a tight-binding model [11]. In both reports, it has been assumed that the dominant contribution to the band comes from phonons close to the KM direction in the Brillouin zone, so-called "*outer*" *processes*. The 2D band of graphite and graphene has been calculated by full integration over the Brillouin zone in the approximation of constant matrix elements [12]. The 2D band of few-layer graphene has been studied within the *ab-initio* approach [13]. The 2D band of SLG, BLG, and triple layer graphene has been calculated within an extended tight-binding model [14]. In the latter two papers, the electron-photon coupling has been considered but the electron-phonon one has been taken wavevector-independent. In all mentioned works, resonant processes with scattering of one electron and one hole (triple resonances) have been omitted.

Recently, the assumption of dominant contribution of "outer" processes has been questioned based on precise calculations of the 2D band of SLG [15,16] and Raman measurement data [17]. It has been argued that, on the contrary, scattering processes, involving phonons along the ΓK direction, so-called "*inner*" *processes*, are dominant. It has also been demonstrated that this band comes primarily from triple resonance processes [15]. The 2D band of BLG has been analyzed assuming dominant "inner" processes within a tight-binding model [18,19] and the *ab-initio* approach [20]. An *ab-initio* calculation of the 2D band intensity with integration over the entire Brillouin zone, accounting for electron-photon and electron-phonon interactions, as well as triple resonances, but restricted to overtone modes, has improved the overall agreement with experiment [21]. 2D' and D + D" bands have also been observed in BLG [6,13] but have not been computed so far. The observation of other two-phonon bands of BLG in the frequency range between 1600 and 2200 $cm^{-1}$ [22-26] and the calculation of the Raman shift of some of these bands within the extended tight-binding model of the band structure and a force constant model of the phonon dispersion [23] have also been reported. The additional two-phonon bands provide a practical way for distinguishing BLG from SLG for future application of BLG in nanoelectronics [22]. Despite the recent advance on the modeling of the Raman shift of the two-phonon bands of BLG, the complete calculation of the intensity of the 2D band and the other, less intense two-phonon Raman bands, which includes both overtone and combination modes, to our knowledge, has not been reported so far.

Here, we calculate the two-phonon Raman bands of BLG within a non-orthogonal tight-binding (NTB) model with model parameters taken over from density functional theory (DFT) studies [27,28]. The electronic band structure [29] and phonon dispersion [30,31], the electron-photon and electron-phonon matrix elements [32], and the electronic linewidth [33] are derived within the NTB model. The computational details are given in Sec. II. The obtained electron and phonon dispersion, and two-phonon Raman bands are discussed in Sec. III. The paper ends up with conclusions (Sec. IV).

## 2. Computational details

Bernal-stacked BLG consists of two graphene layers weakly bound together by Van der Waals forces (Fig. 1). The intralayer and interlayer interactions are described here within the NTB model by means of two different sets of matrix elements of the Hamiltonian and overlap matrix elements, transferred from *ab-initio* studies on carbon dimers [27,28]. The NTB model allows one to obtain the total energy and the forces on the atoms. This feature is indispensable for the atomic structure relaxation, which has to be performed prior phonon dispersion calculations.

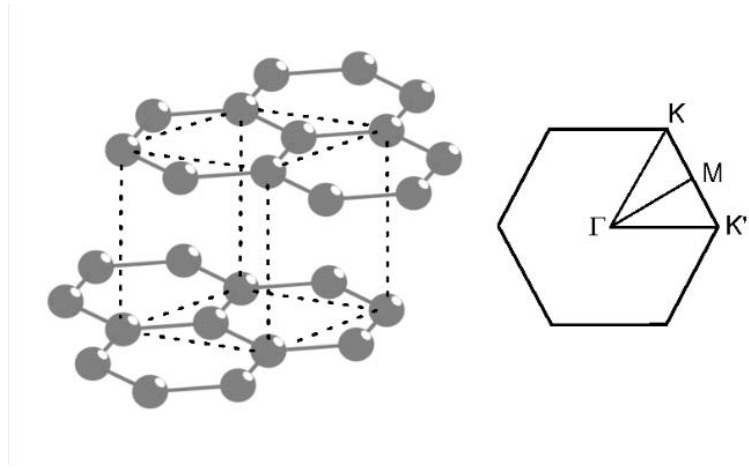

Figure 1. Left: atomic structure of Bernal-stacked BLG. The four-atom unit cell is drawn by dashed lines. Right: Brillouin zone with several high-symmetry points.

The phonon dispersion of BLG is calculated using a dynamical matrix, derived by perturbation theory within the NTB model [30]. The electron-phonon matrix elements, necessary for the calculation of the dynamical matrix, are derived within the same model [32]. The summation over the Brillouin zone in the first-order perturbation term of the dynamical matrix is performed over a $40 \times 40$ mesh of **k** points, for which convergence of the phonon frequencies within 1 cm$^{-1}$ is achieved.

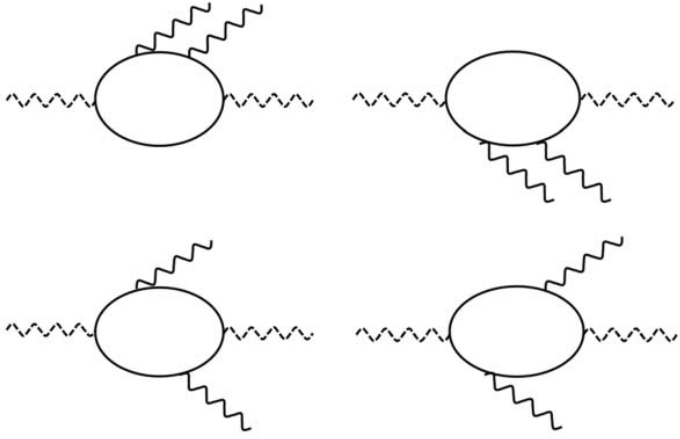

Figure 2. Feynman diagrams for resonant two-phonon Stokes Raman processes. Each process includes incoming and outgoing photons (dashed wavy lines), two phonons with opposite momenta (solid wavy lines), and electrons and holes (fermion loop). The number of different diagrams amounts to eight, accounting for the different chronological order of phonon scattering.

The two-phonon Raman bands of BLG arise from resonant scattering processes. For the considered here Stokes processes, such a process consists of four virtual processes of absorption of a photon with creation of an electron-hole pair, two consecutive processes of scattering of the electron/hole by a phonon, and final recombination of the electron-hole pair with emission of a photon. The momentum is conserved for each virtual process but the energy is conserved only for the entire resonant process. There are altogether eight different types of resonant processes [15,34] (Fig. 2). The two-phonon Raman intensity can be described by fourth-order terms in perturbation theory and is given by the expression [9]

$$I \propto \sum_f \left| \sum_{c,b,a} \frac{M_{fc} M_{cb} M_{ba} M_{ai}}{\Delta E_{ic} \Delta E_{ib} \Delta E_{ia}} \right|^2 \delta(E_i - E_f). \qquad (1)$$

Here, $\Delta E_{iu} = E_i - E_u - i\gamma$, $E_u$ ($u = a, b, c$) are the energies of the intermediate states of the system of photons, electrons, holes, and phonons. $E_i$ is the energy of the initial state with only an incident photon present with energy $E_L$ (laser excitation energy). $E_f$ is the energy of the final state with a scattered photon and two created phonons. $M_{uv}$ are the matrix elements for virtual processes between initial, intermediate, and final states. In particular, $M_{ai}$ and $M_{fc}$ are the momentum matrix elements for creation and recombination of an electron-hole pair, respectively. $M_{ba}$ and $M_{cb}$ are the electron/hole-phonon matrix elements for scattering between intermediate states. The electron-photon and electron-phonon matrix elements are calculated explicitly within the NTB model [32]. $\gamma$ is the sum of the halfwidths of pairs of electron and hole states [33]. It is taken twice larger than for SLG [16] in order to account for both the electron-phonon and electron-electron scattering contributions [21], namely, $\gamma = 25.2 E_L + 6.9 E_L^2$ ($\gamma$ is in meV and $E_L$ is in eV). The Dirac delta function ensures energy conservation for the entire resonant process. In the calculations, it is replaced by a Lorentzian with a halfwidth of 5 cm$^{-1}$. The summation over the intermediate states runs over the four valence and conduction bands, and over all electron wavevectors **k**. The summation over the final states runs over all pairs of phonon branches and phonon wavevectors **q**. For both summations, convergence is reached with a 400 × 400 mesh of **k** and **q** points in the Brillouin zone. Both "inner" and "outer" processes are explicitly included by the summations over the entire Brillouin zone. All the eight different types of resonant processes are included in the calculation of the Raman intensity. We restrict ourselves to parallel scattering configuration and backscattering geometry with respect to the graphene layers. The *quantum interference* due to the virtual processes is explicitly accounted for by Eq. (1).

## 3. Results and Discussion

*3.1 Electronic band structure, phonon dispersion, and two-phonon Raman bands*

The Raman intensity is resonantly enhanced whenever the laser excitation energy matches an optical transition of the studied material. In SLG, this can happen at any laser excitation energy, used in the Raman experiments, because the valence and conduction bands have the form of Dirac cones with common apices near the Fermi level. Since this can take place for any intermediate state, there can be single-, double-, and triple-resonant processes, all of them described by Eq. (1). We also note that, while the resonant scattering in SLG has long been termed "double-resonant", this term does not correctly describe the dominant contribution to the Raman intensity, which comes from triple-resonant processes [15].

In BLG, the electronic bands double in number, compared to SLG, because of the doubled unit cell. The interlayer interactions lift the band degeneracy and band splitting appears. The Dirac cones of the SLG are replaced by pairs of parabolic bands denoted as $\pi_1$ and $\pi_2$, and $\pi_1^*$ and $\pi_2^*$, which are also commonly referred to as the Dirac cones (Fig. 3). Hence, the Raman scattering in BLG is resonant, similarly to SLG. The analysis of the selection rules has allowed to conclude that, along the high-symmetry direction ΓKMKΓ, only electron-hole processes $\pi_1 \leftrightarrow \pi_1^*$ and $\pi_2 \leftrightarrow \pi_2^*$, further on denoted by "1" and "2", are allowed [11]. Although all electron-hole processes are allowed away from this direction, these two processes are dominant, while the other processes can be neglected. Nevertheless, in the calculations, we consider all electron-hole processes but discuss the Raman bands only in terms of the contributions from these two processes.

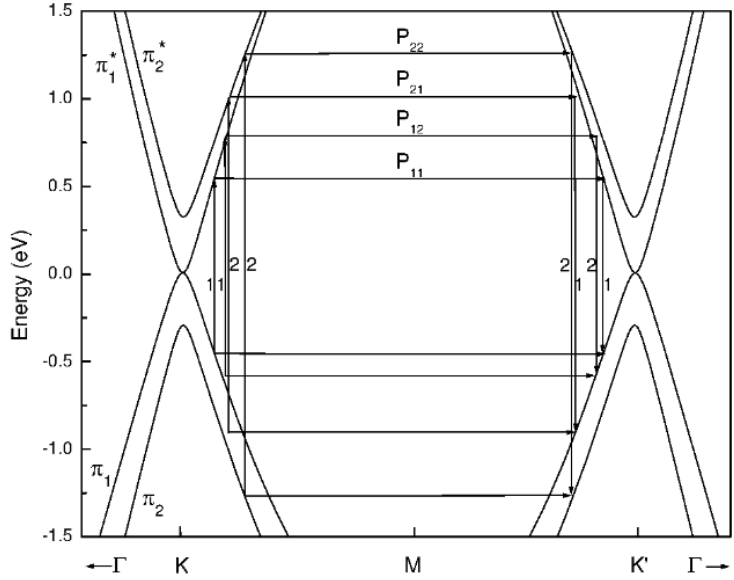

Figure 3. Electronic band structure of Bernal-stacked BLG along the ΓKMK'Γ direction of the Brillouin zone. The Fermi energy is chosen as zero. The vertical lines represent schematically the dominant electron-hole creation/annihilation processes "1" and "2". The horizontal lines show schematically intervalley scattering processes for the particular case of "inner" processes. The polygons of arrows are the four resonant processes with intervalley scattering: $P_{11}$, $P_{12}$, $P_{21}$, and $P_{22}$. There are altogether eight such sets of intervalley processes. There are also eight corresponding sets of intravalley processes (not shown).

A resonant process consists of scattering of the electrons and holes between two Dirac cones at the K and K' points (intervalley scattering) or within the same Dirac cone (intravalley scattering), and final annihilation of the electron-hole pairs. In SLG, there are eight types of resonant processes for each of the two types of scattering, while the doubling of the bands in BLG increases four times the number of the resonant processes. Using the notation of the electron-hole processes "1" and "2", the four resonant processes for each of the eight types of resonant processes can be written as $P_{11}$, $P_{12}$, $P_{21}$, and $P_{22}$. These processes are drawn by closed polygons of arrows for one of the eight possible types of resonant processes with intervalley scattering in Fig. 3.

BLG belongs to space group #164 (P−3m1) [7]. It has a unit cell consisting of four carbon atoms (Fig. 1). The twelve Γ-point vibrational modes $2A_{1g} + 2E_g + 2A_{2u} + 2E_u$ are classified as Raman-active modes $2A_{1g} + 2E_g$, infrared-active modes $A_{2u} + E_u$, and acoustic modes $A_{2u} + E_u$. The "g" ("u") modes are symmetric (antisymmetric) with respect to inversion. The nondegenerate modes $A_{1g}$ are symmetric out-of-plane motions of the two graphene layers with in-phase and out-of-phase displacement of the atoms of each layer (ZO' and $ZO^+$ modes, respectively). The doubly degenerate modes $E_g$ are symmetric in-plane motions of the two layers with in-phase and out-of-phase atomic displacement (LO' + TO' and $LO^+ + TO^+$ modes, respectively). The $A_{2u}$ modes are antisymmetric out-of-plane motions of the layers with in-phase and out-of-phase atomic displacement (ZA and $ZO^−$, respectively). Finally, the $E_u$ modes are antisymmetric in-plane motions of the two layers with in-phase and out-of-phase atomic displacement (LA + TA and $LO^− + TO^−$ modes, respectively). In the literature, a simplified notation of the BLG modes is also used, which does not make difference between symmetric and antisymmetric modes and each pairs of such modes is denoted by a single symbol: ($LO^−$, $LO^+$) → LO, ($TO^−$, $TO^+$) → TO or iTO, ($ZO^−$, $ZO^+$) → ZO or oTO, (LO', LA) → LA, and (TO', TA) → TA. The phonon branches will be denoted by the symbols of the Γ phonons, except for the branches close to the KM direction, which normally have small contributions to the Raman bands.

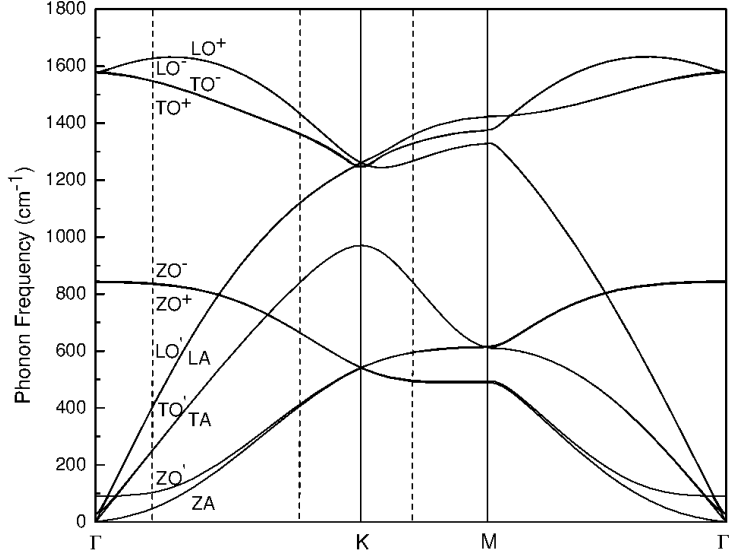

Figure 4. Phonon dispersion of Bernal-stacked BLG along high-symmetry directions in the Brillouin zone. The phonon branches are denoted by the symbols of the corresponding Γ phonons. The three vertical dashed lines, from left to right, indicate approximately the phonons for intravalley scattering, "inner" and "outer" intervalley scattering processes, respectively, at $E_L$ = 2.33 eV.

The calculated phonon dispersion of BLG is shown in Fig. 4. For better agreement of the obtained phonon dispersion with experiment, the frequency of the in-plane phonons is downscaled by a factor of 0.9 in order to remedy the predicted overestimation of the frequency, which is inherent to tight-binding models [16,31]. The comparison of the phonon dispersion with available experimental data has been discussed previously [10,16,31]. The order of the phonon branches is established by careful inspection of the calculated phonon eigenvectors. In most cases, the symmetric branch has either higher or lower frequency than the antisymmetric partner. This is not the case with the $LO^{\pm}$ branches, which intersect at roughly one sixth of the distance between the zone center and zone boundary, beyond which branch $LO^+$ has higher frequency than branch $LO^-$ (see, also, Ref. [35]).

The two-phonon Raman bands of BLG are due to overtone and combination modes, each of which involves two phonons with opposite momenta near the Γ, K, and M points of the Brillouin zone. Such phonons will be referred to as the Γ, K, and M phonons, respectively. We adopt the notation of the two-phonon modes of the form XY@Z, where X and Y are the symbols of the participating phonons, phonon X having higher frequency than phonon Y; Z is a Brillouin zone point. This notation describes most precisely the modes and corresponds to that already in use by many groups.

The selection rules for scattering processes along the ΓKMKΓ direction impose restrictions on the participating phonons. Thus, as far as the TO and LO branches are concerned, electron scattering $\pi_1^* \leftrightarrow \pi_1^*$ and $\pi_2^* \leftrightarrow \pi_2^*$ is allowed for TO phonons, while electron scattering $\pi_1^* \leftrightarrow \pi_2^*$ is allowed for LO phonons [36]. Away from this direction, these restrictions are relaxed and all combinations of electronic transitions and pairs of phonons should be considered and are considered in the calculations here.

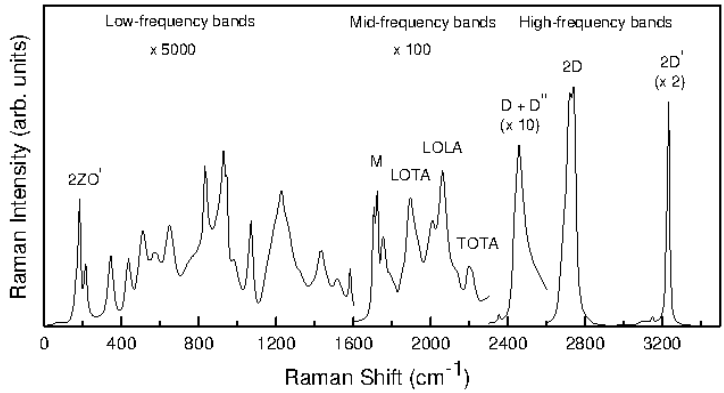

Figure 5. Two-phonon Raman bands of BLG at $E_L$ = 2.33 eV. The spectrum is divided into three regions with intensity of the bands, differing by a few orders of magnitude. Several experimentally observed Raman bands are indicated.

The two-phonon Raman spectrum of BLG at $E_L$ = 2.33 eV, shown in Fig. 5, can be divided into three regions according to the intensity of the characteristic bands, namely, low-, mid-, and high-frequency regions with intensity ratio 1:50:5000. In principle, each pair of phonon branches gives rise to an overtone or a combination mode. It can happen that such modes partly overlap and form an isolated Raman band. The latter will be assigned to the most intense mode, while the low-intensity ones, though contributing to the isolated feature, will not be shown and discussed. Isolated low-intensity bands will be omitted from the discussion as well. The bands will be denoted by the symbols of the most intense modes. However, for conformity with existing notations, the latter symbols will also be used in a simplified form, without designating the parity of some of the phonons, e.g., LOLA, TOTA, etc., as well as the widely accepted concise notation with a single letter, e.g., D, D', etc.

In SLG, the eight types of resonant processes give rise to closely positioned peaks, which are observed as a single symmetric Raman band [15,16]. In BLG, due to the doubling of the Dirac cones, the number of resonant processes increases four times and the Raman band shape becomes more complicated. Indeed, the four resonant processes $P_{11}$, $P_{12}$, $P_{21}$, and $P_{22}$ are associated with phonons with different wavevectors. For a given laser excitation energy, there are only three distinct phonon wavevectors, related by inequalities, e.g., $|\mathbf{q}_{11}| < |\mathbf{q}_{12}| \approx |\mathbf{q}_{21}| < |\mathbf{q}_{22}|$ for K phonons and dominant "inner" processes [21] and $|\mathbf{q}_{11}| > |\mathbf{q}_{12}| \approx |\mathbf{q}_{21}| > |\mathbf{q}_{22}|$ for $\Gamma$ phonons (Fig. 3). Then, for increasing sum of the frequencies of the two phonons with increasing wavevector, the three wavevectors will give rise to three bands at positions $\omega$, related by the inequalities $\omega_{11} < \omega_{12} \approx \omega_{21} < \omega_{22}$ for K phonons and $\omega_{11} > \omega_{12} \approx \omega_{21} > \omega_{22}$ for $\Gamma$ phonons, etc. Depending on the electron-phonon matrix elements [11], the four processes will produce a single-peak, double-peak, or triple-peak Raman feature.

In the next subsections, we present and discuss the results of the complete calculations of the two-phonon Raman bands of BLG.

3.2 *Low-frequency two-phonon Raman bands*

The low-frequency region below 1600 cm$^{-1}$ contains a number of very-low intensity single-peak and double-peak two-phonon modes (Fig. 6). The $\Gamma$ phonons give rise to the double-peak modes 2ZO', TO'ZO', 2TA, LO'ZO', 2LO', TO'ZO$^+$, and LO'ZO$^+$, and the single-peak ones 2TO', LATA, ZO$^+$ZO', 2LA, and TO$^-$ZA (Fig. 6(a)). There are three prominent K phonon modes (Fig. 6(b)): two single-peak modes ZO$^\pm$ZO' and a double-peak one LAZO'. There are also four modes due to the divergent phonon density of states of the pairs of branches (ZO', LA) and (ZO$^+$, ZO$^-$) at the M point, namely, 2ZA, 2ZO', and 2ZO$^\pm$.

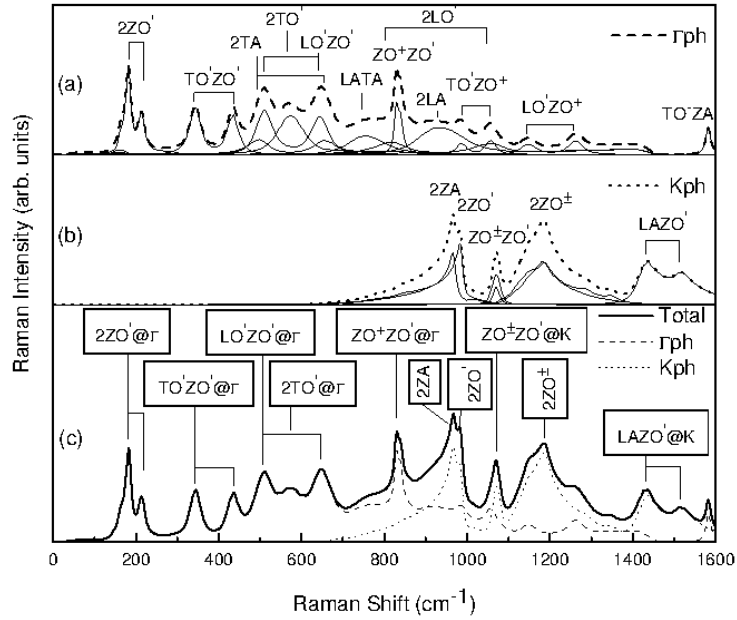

Figure 6. Low-frequency two-phonon Raman bands of BLG at $E_L = 2.33$ eV. (a) Γ phonon modes, (b) K phonon modes, and (c) Total Raman spectrum.

The Raman spectrum, formed by the superimposed two-phonon modes, exhibits fewer bands, as shown in Fig. 6(c), namely, 2ZO'@Γ, TO'ZO'@Γ, LO'ZO'@Γ, 2TO'@Γ, ZO⁺ZO'@Γ, 2ZA@M, 2ZO'@M, ZO±ZO'@K, 2ZO±@M, and LAZO'@K. The experimental work on the two-phonon bands in this frequency range is limited to the study of the double-peak band in the frequency range 145 - 220 cm$^{-1}$ [37]. It has been assigned to the 2ZO'@Γ mode and, especially, to resonant processes $P_{22}$ and $P_{11}$, which is confirmed here. The measured low-frequency bands can be used for reconstruction of the phonon dispersion near the Γ, K, and M points of the Brillouin zone.

The dependence of the low-frequency modes on $E_L$ is shown in Fig. 7 in comparison with a few measured data points. Most of the bands have a distinct dependence on $E_L$, i.e., they are dispersive. Although this dependence is only approximately linear, it is described here by the slope of the linear interpolation of the calculated curves, which will be addressed to as the *dispersion rate*. The derived dispersion rates, in order of increasing frequency, are 41 and 66 cm$^{-1}$/eV (2ZO'@Γ), 156 and 169 cm$^{-1}$/eV (TO'ZO'@Γ), 251 and 249 cm$^{-1}$/eV (LO'ZO'@Γ), 262 (2TO'@Γ), −206 and −197 cm$^{-1}$/eV (LAZO'@K). Bands ZO⁺ZO'@Γ and ZO±ZO'@K are almost dispersionless and bands 2ZA@M, 2ZO'@M, 2ZO±@M are non-dispersive.

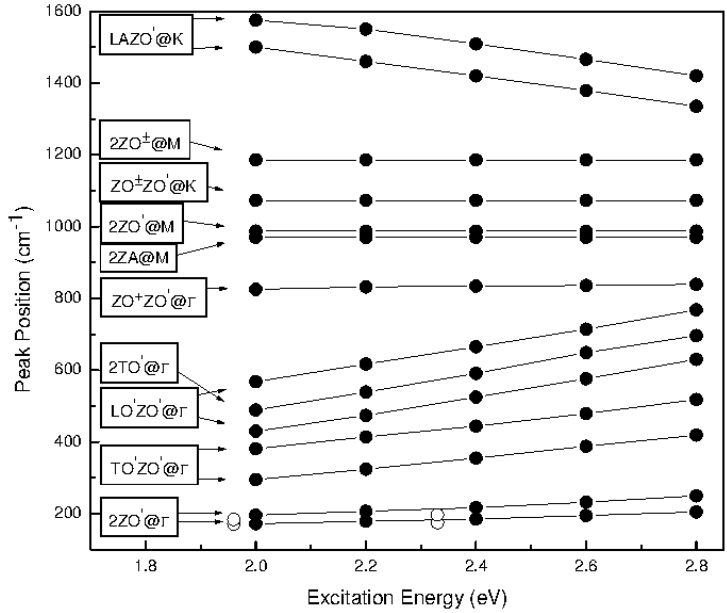

Figure 7. Low-frequency two-phonon Raman bands of BLG at several $E_L$: calculations (solid symbols) and experimental data [37] (open symbols). The lines are guides to the eye.

3.3 *Mid-frequency two-phonon Raman bands*

The calculated mid-frequency two-phonon Raman spectrum of BLG between 1600 and 2300 cm$^{-1}$ at $E_L$ = 2.33 eV has a number of bands, due to combination modes, as shown in Fig. 8. The region between 1600 and 1800 cm$^{-1}$ exhibits the modes LO$^-$ZO'@Γ (Fig. 8(a)) and TO$^+$ZO'@K (Fig. 8(b)). In this frequency region, two bands have been observed in the Raman spectra of single-walled carbon nanotubes [38] and few-layer graphene [22]. In the latter paper, the two bands have been denoted as M$^-$ and M$^+$, referred to as the *M band*, and assigned to the 2ZO mode. Since the ZO phonons do not contribute to the two-phonon Raman spectra in graphene, their activation in nanotubes has been attributed to curvature effects. Similarly, the activation of these phonons in BLG has been explained with the interlayer interactions. In recent studies of the phonons ZO'@Γ of few-layer graphene [23-25], the observed band M$^-$ has been assigned to mode LOZO'@Γ, while M$^+$ has been attributed to mode 2ZO [23,25]. Based on our results, we confirm the former, but reject the latter assignment. The absence of an observable 2ZO mode can be justified from a theoretical point of view with the small electron-phonon matrix element for the ZO phonon, which is present in the expression of the Raman intensity, Eq. (1), raised on fourth degree. The calculated intensity of mode 2ZO@K is found here to be ≈300 times smaller than that of mode TO$^+$ZO'@K and, therefore, the former band should hardly be observable. As we shall see, the modes in the mid-frequency range normally include one phonon with large electron-phonon coupling, such as LO@Γ or TO@K. Another argument from our calculations for the assignment of the observed bands M$^\pm$ is their dependence on $E_L$. The experiment clearly shows [26] that the M band splitting decreases with increasing $E_L$ and such behavior is reproduced here, as shown in Fig. 9. Therefore, *the observed band* M$^+$ *should be assigned to mode* TO$^+$ZO'@K *rather than to mode* 2ZO@Γ. Since the M band is only observed in BLG but not in SLG, it can be used for distinguishing between the two structures.

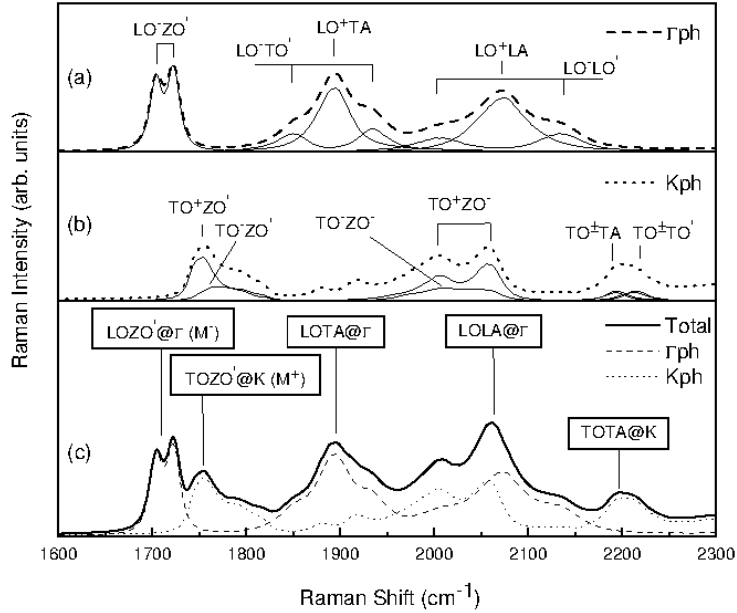

Figure 8. Mid-frequency two-phonon Raman bands of BLG at $E_L$ = 2.33 eV. (a) Γ phonon modes, (b) K phonon modes, and (c) Total Raman spectrum.

At about 1900 cm$^{-1}$, the single-peak mode LO$^-$TA@Γ and the double-peak one LO$^-$TO'@Γ are predicted. The two modes correspond to an observed pair of bands in BLG, which has been previously assigned to modes LOTA@Γ and TOTA@Γ [23] or LOTA@Γ [22,26]. The corresponding band in SLG is a sharper and narrower single-peak one, and has been assigned to mode LOTA@Γ.

At about 2040 cm$^{-1}$, there are two comparable contributions from Γ and K phonons. The single-peak mode LO$^+$LA@Γ and the double-peak one LO$^-$LO'@Γ give rise to band, denoted by LOLA@Γ. The modes TO$^+$ZO$^−$@K and TO$^−$ZO$^−$@K bring forth a band, denoted briefly by TOZO@K. The latter band does not exist in SLG, because of the negligible electron-phonon coupling of phonon ZO$^−$@K. However, in BLG, this phonon is activated for electron scattering due to the interlayer interactions. The Raman band, located at this position in SLG, has been assigned to mode LOLA@Γ and that in BLG has been assigned to modes TOLA@Γ and LOLA@Γ [23] or mode LOLA@Γ [22]. However, according to our calculations, *the observed band should be assigned to both overlapping modes LOLA@Γ and TOZO@K*.

Finally, at about 2160 cm$^{-1}$, a single-peak band, due to the four modes TO$^\pm$TA@K and TO$^\pm$TO'@K and denoted by TOTA@K, is predicted. The observed Raman band in this frequency region has been previously assigned to band TOTA@K [26], which is confirmed here.

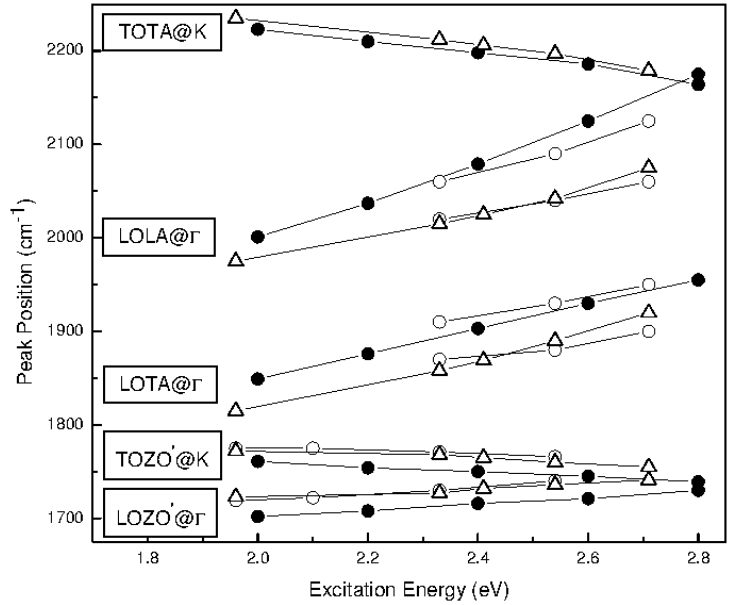

Figure 9. Mid-frequency two-phonon Raman bands of BLG at several different $E_L$: calculations (solid symbols) and experimental data (open circles [23] and open triangles [26]). The lines are guides to the eye.

The derived dependence of the mid-frequency bands on $E_L$ (Fig. 9) is characterized by the dispersion rates: 34 cm$^{-1}$/eV (LOZO'@Γ), −27 cm$^{-1}$/eV (TOZO'@K), 133 cm$^{-1}$/eV (LOTA@Γ), 218 cm$^{-1}$/eV (LOLA@Γ), and −71 cm$^{-1}$/eV (TOTA@K). These values compare well to the experimental ones [22,23,26] except for band LOLA@Γ, where the dispersion rate is overestimated here by about 30%.

## 3.4 *High-frequency two-phonon Raman bands*

The high-frequency two-phonon Raman spectrum from 2300 to 3300 cm$^{-1}$, shown in Fig. 10, exhibits three intense bands, known as TOLA@K (or D + D"), 2TO@K (or 2D), and 2LO@Γ (or 2D'). The lowest-frequency band TOLA@K in this frequency range is entirely composed of combination modes, namely, TO$^+$LA, TO$^-$LO', TO$^+$LO', and TO$^-$LA, which are due to intervalley scattering by K phonons (Fig. 10). The resulting shape of the TOLA@K band is asymmetric with a steep left shoulder and a broad right shoulder. Although the observation of this band has often been reported (e.g., [7,8]) as far as we know, no calculations of this band have been published.

The most intense band 2TO@K originates from intervalley scattering by K phonons. It has been studied in a number of works [6,11,13,14,18-21] and several calculations have also been reported [11,18,20,21]. The relative intensity of this band with regard to the first-order G band is smaller and its width is larger than in SLG. The complex shape of this band has been fitted by four peaks for the four resonant processes [11] (Fig. 10). Recent calculations, restricted to overtone modes, have proved that the contributions of the P$_{12}$ and P$_{21}$ processes are closely positioned and should be fitted by a single peak, thus reducing the number of the fitting peaks to three [21].

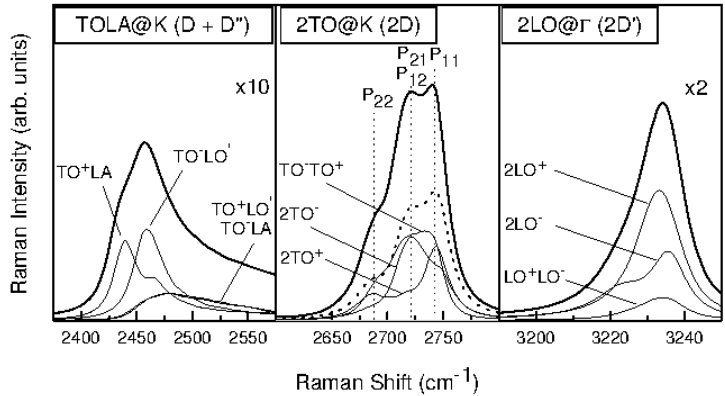

Figure 10. High-frequency two-phonon Raman bands of BLG at $E_L$ = 2.33 eV: TOLA@K (D + D"), 2TO@K (2D), and 2LO@Γ (2D') bands. The thin curves are the major contributions to the bands. Middle panel: The dotted curve is the contribution of overtone modes only. The vertical dotted lines indicate approximately the position of the peaks due to P$_{11}$, P$_{12}$, P$_{21}$, and P$_{22}$ resonant processes.

We extend the latter calculations of the 2TO@K band to account for not only the overtone modes, but also for the combination ones. The obtained band at $E_L$ = 2.33 eV consists of overtone and combination modes, originating from the two K phonons TO¯ and TO⁺ and the four resonant processes (Fig. 10). The 2TO¯ mode, originating primarily from $P_{12}$ and $P_{21}$ processes, is positioned closely to the center of the 2TO@K band. The 2TO⁺ mode comes from $P_{11}$ and $P_{22}$ processes, involving phonons with wavevectors, which are smaller and larger than are those for the $P_{12}$ and $P_{21}$ processes. The resulting band has two peaks: the lower-frequency one is due to $P_{22}$ processes and the higher-frequency one is due to $P_{11}$ processes. Apart from the two overtone modes, a significant contribution comes from the combination mode TO¯TO⁺. Although the latter does not produce additional features of the 2TO@K band, it brings about a visual modification of the band shape. The predicted 2D band shape corresponds fairly well to the observed one [6,11,13,14,18-21]. However, the intensity of our 2D band is possibly lower than the experimental one by a factor of 2, which can be corrected by explicitly considering GW corrections to the phonon dispersion, as it has been commented in the case of SLG [16].

Previous studies of the 2TO@K band splitting have been based on the assumption of dominant "outer" processes [6,11,14] and have predicted the same relative positions of the contributions of the $P_{11}$ and $P_{22}$ processes as here. The importance of the "inner" processes has been underlined in a subsequent work [18], which has been supported by extensive calculations [19] but the positions of the contributions have been erroneously reversed attributing the lower-frequency peak to $P_{11}$ processes and the higher-frequency one to $P_{22}$ processes. Such an incorrect assignment is present in another paper as well [20]. The correct assignment of the peaks of band 2TO@K to resonant processes has been given in a recent work [21] and confirmed by our results.

The peak separations contain information about the phonon branch splitting [21]. The unequal splitting between the three peaks can be explained by the doubling and splitting of the TO branch in BLG. Denoting the Raman shift of the peaks in order of increasing frequency by $\omega_1$, $\omega_2$, and $\omega_3$, the vertical TO branch splitting at $E_L$ = 2.33 eV, determined from the expression $\Delta\omega = (\omega_2 - (\omega_1 + \omega_3)/2)/2$, is found to be about 4 cm⁻¹. This value is twice larger than a previous DFT result [6], but is a few times smaller than the GW-corrected DFT value, while all of them underestimating the experimental data. The derived outmost peak separation $\omega_3 - \omega_1$ = 61 cm⁻¹ slightly underestimates the experimental data [11]. Previous parameter-free calculations either significantly underestimate [14], or overestimate this separation [21]. Obviously, additional efforts are needed for the more accurate prediction of the shape of the 2TO@K band.

The 2LO@Γ band is the highest-frequency two-phonon one, predicted in the Raman spectra of BLG (Fig. 10). Our calculations show that it is formed primarily from the overtone modes $2LO^-$ and $2LO^+$, which arise from intravalley scattering by the Γ phonons $LO^-$ and $LO^+$. Unlike band 2TO@K, here there is only a minor contribution from the combination mode $LO^+LO^-$. Similarly to the case of band 2TO@K, the unequal splitting between the peaks of the overtone modes contains information about the LO branch splitting close to the Γ point. The derived splitting at $E_L = 2.33$ eV of about 2 cm$^{-1}$ is three times smaller than the GW-corrected DFT value [21] but this underestimation is unlikely to bring about significant modification of the band shape. This band is normally observed as an intense symmetric feature in the Raman spectra [7,8]. To our knowledge, no calculations of this band have been reported.

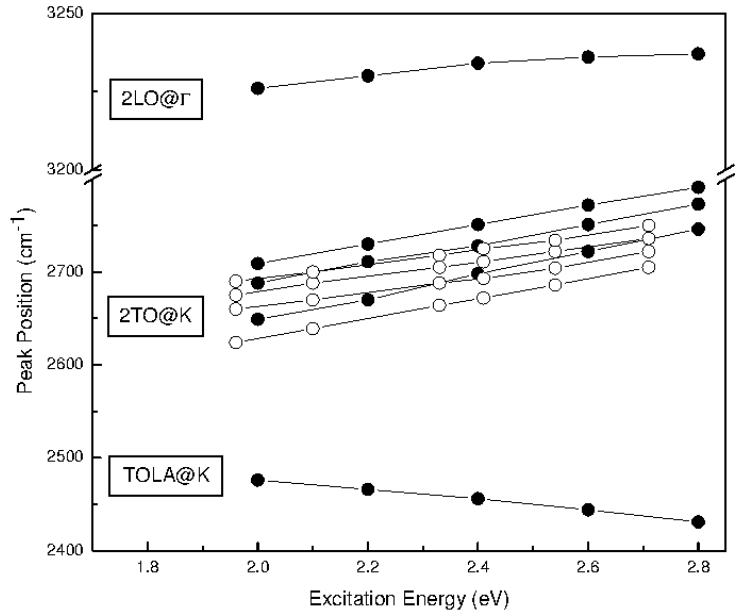

Figure 11. High-frequency two-phonon Raman bands of BLG at several different $E_L$: calculations (solid symbols) and experimental data [11] (open symbols). The lines are guides to the eye.

The three bands in the high-frequency Raman spectrum are dispersive, as shown in Fig. 11. The derived dispersion rates are −56 cm$^{-1}$/eV (TOLA@K), 98 cm$^{-1}$/eV (2TO@K), and 14 cm$^{-1}$/eV

(2LO@ Γ). These values agree well with the experimental and theoretical ones for SLG [16], as well as with the value of ≈89 cm$^{-1}$/eV for BLG [11]. The change of the shape of the 2TO@K band with increasing $E_L$ corresponds visually to the observed behavior for BLG [6,14,20,21].

## 4. Conclusions

We have presented results of the complete calculation of the two-phonon Raman bands of graphene within a non-orthogonal tight-binding model with DFT-derived matrix elements. We have shown that the shape of these bands is mostly due to overtone and combination modes of phonons, close to the Γ and K points in the Brillouin zone, and have assigned all major observed Raman bands to two-phonon modes. In particular, we have argued that the previously reported assignment of the Raman band M$^+$ is incorrect and have proposed an assignment, based on our results. We have identified another major contribution to the Raman band, previously assigned only to mode LOLA. We have calculated the laser excitation dependence of the Raman bands, which can be used for assignment of the observed Raman features of BLG to overtone and combination modes. The proposed model can be utilized for similar calculations on few-layer graphene with various layer stacking.

## Acknowledgments

V.N.P. acknowledges financial support from EU Seventh Framework Programme project INERA under grant agreement number 316309.